\documentclass[aps,prl,twocolumn,showpacs]{revtex4}
\usepackage{amsmath}
\usepackage{subfigure}
\usepackage{graphicx}
\usepackage{epsfig}
\usepackage{mathdots}

\usepackage{epsf}

\newlength{\upit}\upit=0.1truein

\newcommand{\ltappr}{{{\lower4pt\hbox{$<$} } \atop \widetilde{ \ \ \ }}}
\newlength{\bxwidth}\bxwidth=1.5 truein

\newcommand{\si}{\sigma}
\newcommand{\rarrow}{\rightarrow}
\newcommand{\bk}{{\bf k}}
\newlength{\figwidth}
\newlength{\shift}
\usepackage{epsf}

\newcommand \bea {\begin{eqnarray} }
\newcommand \eea {\end{eqnarray}}

\newcommand{\beg}{\begin{equation}}
\newcommand{\en}{\end{equation}}

\newcommand{\eps}{\varepsilon}

\begin{document}

%%% abbriviations %%%
\newcommand{\dg}{^{\dagger }}
\newcommand{\vk}{\mathbf k}
\newcommand{\vp}{\mathbf p}
\newcommand{\bfr}{\mathbf r}
\newcommand{\veps}{\varepsilon}
%%%%%%%%%%%%%

\title{Electron Cotunneling into a Kondo Lattice}

\date{\today}

\author{Marianna Maltseva, M. Dzero and P. Coleman}
\affiliation{Center for Materials Theory, Rutgers University, Piscataway, New
Jersey, 08854, USA}

\begin{abstract}
Motivated by recent experimental interest in tunneling into heavy electron materials,
we present a theory for electron tunneling into a Kondo lattice.
The passage of an electron into a Kondo lattice is accompanied by a simultaneous
spin flip of the localized moments via cotunneling mechanism. We compute the tunneling
current with the large-$N$ mean field theory. In the absence of disorder, differential tunneling 
conductance exhibits two peaks separated by the hybridization gap. 
Disorder effects lead to the smearing of the gap resulting in a Fano lineshape.
\end{abstract}


\maketitle

% Kondo lattice materials are  inhomogeneous in both real and momentum space, so that  their investigation  calls for ultra-fine experimental tools.
%Discoveries  are often aided by advances in experimental techniques. 
Major developments in scanning tunneling electron
spectroscopy (STEM) over the last decade, particularly  as a probe of cuprate
superconductors  \cite{ibm,kapitulnik,sdavis,Hanaguri2009}, suggest that 
this tool will find increasing utility as an atomic-scale 
probe of many-body phenomena
in  new classes of materials. One area of 
particular promise lies in the application of STEM 
to heavy fermion materials. 

Heavy fermion compounds contain a dense lattice of localized magnetic
moments interacting with a sea of conduction electrons to form a
``Kondo lattice''  \cite{Sarrao2007,Coleman2008}.  These materials
exhibit  a diversity of many body behaviors, including anisotropic superconductivity,
Kondo insulating behavior and quantum criticality. Motivated by recent tunneling 
experiments on $f$-electron materials \cite{Davis2009,Park2008,Yazdani2009},  
in this paper we develop a theory for tunneling into a coherent Kondo lattice. 

How do electrons tunnel into a Kondo lattice, where the main degrees
of freedom are local moments? Since direct tunneling into localized magnetic orbitals is 
blocked by Coulomb interactions, the naive expectation is that the
electrons can only tunnel into the surrounding conduction sea. 
In 1960s Anderson and Appelbaum \cite{wyatt64,Appelbaum1966,Anderson1966}
recognized that magnetic ions actively participate in the
tunneling process via a ``cotunneling mechanism'' \cite{Averin1990,glazman}
in which the passage of a tip electron into the conduction sea occurs
cooperatively with a spin-flip of localized moments. The manifestation of 
cotunneling in the tunneling conductance of quantum dots and magnetic atoms 
adsorbed on surfaces is well established experimentally 
\cite{Averin1990,glazman,goldhabergordon,vankempen}.
Here we examine the effect of these processes on tunneling into
a coherent band of excitations of a Kondo lattice, deriving a new
expression for the tunneling current into a Kondo lattice in terms 
of the Green's function of composite co-tunneling operators. 
Using the large-$N$ approximation, we show how cotunneling processes open a 
direct tunneling channel between the tip and the composite quasiparticle states of the
Kondo lattice. Once coherence develops, cotunneling and direct tunneling processes 
interfere,  giving rise to distinctive two peak structures in tunneling spectra. 

\begin{figure}[h]
\includegraphics[width=3.2in]{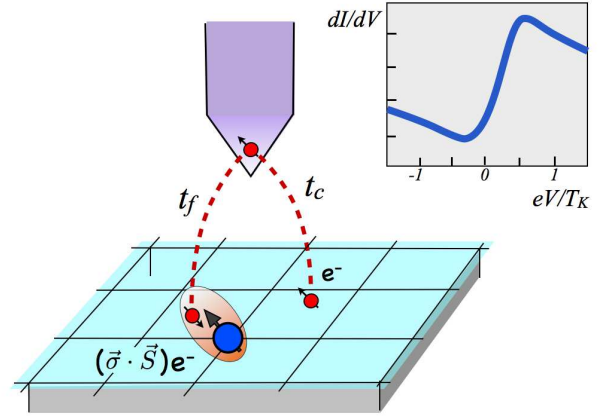}
\caption{(Color online) Electron tunneling into a heavy-fermion
material involves two parallel processes: direct tunneling with 
amplitude $t_c$ into the conduction sea and 
cotunneling with amplitude $t_f$
into a composite combination of the conduction electron and local
magnetic $f$-moments.  
These composite states are expected to develop coherence below 
the Kondo temperature $T_K$. Inset shows a typical differential
conductance curve observed for tunneling into a single Kondo ion.}
\end{figure}

%In this work, we develop a theory of scanning tunneling spectroscopy
%in a Kondo lattice. We find that  
%tunneling into a Kondo lattice develops
%two components: direct tunneling from the tip into the conduction sea
%and ``cotunneling''  
%which involves the spins of the Kondo lattice.
%At low energies, the cotunneling processes mediate weakly hybridize 
%the tip electrons with the composite Landau quasiparticles 
%of the Kondo lattice. 
%The interference between the direct tunneling and the cotunneling channels
% leads to a novel asymmetric lineshape, which
%has two peaks and a gap. 
%
%The presence of the peaks suggests that 
%the interference is more dramatic in the case of Kondo lattice than in
%the single impurity case, because of the coherence.
%These features should be observed in future tunneling experiments on Kondo lattice materials.
%

We begin by writing down the Kondo lattice Hamiltonian in the presence
of a tunneling probe, which takes the form
$\hat{H}=\hat{H}_{KL}+\hat{H}_{tip}+\hat{H}_T$,
where 
\beg\label{eq1}
\hat{H}_{KL}=\sum_{\bk,\sigma}\epsilon_\bk c_{\bk\sigma}\dg c_{\bk\sigma}+
{J}\sum_{j}
\vec{S}_f(j)\cdot (c\dg_{j\alpha }\vec{\sigma }_{\alpha \beta }c_{j\beta } )
\en 
is the unperturbed Kondo lattice Hamiltonian,  
$c_{j\sigma}=
\frac{1}{\sqrt{V}}\sum_{\bk} c_{\bk \sigma }e^{i{\bf k}\cdot {\bf R}_{j}}$
creates a conduction electron and $\vec{S}_{f} (j)$ is the
spin operator of a localized $f$-electron at site $j$, respectively.
The term
$\hat{H}_{tip}=\sum_{\bk\sigma}\epsilon_{\bk}\hat{p}_{\bk\sigma}\dg\hat{p}_{\bk\sigma}$
describes the electrons in the tip. 
The crucial new feature of this model lies in the composite character
of the tunneling Hamiltonian.  
When the tip lies in the vicinity of site $0$, the tunneling  Hamiltonian is given by  
\begin{equation}\label{tun}
\hat{H}_{T}=\hat{p}_{0\alpha}\dg\psi_{0\alpha } + {\rm H. c.},
\end{equation}
where 
\begin{eqnarray}\label{eq2}
\psi_{0\alpha }=  t_{c}\hat{c}_{0\alpha}+ \tilde{t}_{f}
\left({\vec \sigma}_{\alpha\beta}\cdot{\vec S}_f(0)\right)\hat{c}_{0\beta}
\end{eqnarray}
contains a direct tunneling term of amplitude
$t_{c}$ and a ``cotunneling term'' of amplitude ${t}_{f}$.
From the equations of motion, the tunneling current operator is
\begin{equation}\label{}
\hat I =  e \dot{N}_{c}= \frac{ie}{\hbar }\sum_{\alpha}\left(\psi \dg_{0\alpha }p_{0 \alpha }- {\rm H.c.} \right),
\end{equation}
where $N_{c} = \sum_{\bk\si}c\dg_{\bk\si}c_{\bk\si} $
is the number operator of the conduction electrons. From the form of
$\hat I$  and $\hat H_{T}$, we see that the passage of an electron from
the tip into the lattice is accompanied by a spin-flip of a local
moment. In this way, one particle states in the tip are coupled to the
composite fields which define heavy electron quasiparticles. 

The Hamiltonian (\ref{tun}) is a Kondo lattice generalization of 
the Anderson-Appelbaum tunneling
Hamiltonian \cite{Appelbaum1966,Anderson1966}, first
introduced to explain
zero-bias anomalies associated with tunneling between two metallic leads 
via a single localized moment.
Similar models have subsequently
been used to describe tunneling through a quantum dot \cite{glazman}.
The cotunneling component of $H_{T}$ can be understood as a
result of mixing between states in the tunneling tip
and the localized orbitals of the Kondo lattice. This process
distorts the  symmetry  of the Wannier states that hybridize 
with the localized moments, partially delocalizing them into the tip.
A derivation of the cotunneling terms can be done by carrying out 
a Schrieffer-Wolff transformation on the Anderson model describing
the lattice and the tip \cite{Schrieffer1966,Anderson1966,glazman}.
In the Anderson model, the localized $f$-electrons hybridize with the
conduction electrons. When a tip is introduced above 
site $0$ of the lattice, tunneling between the $f$-state and the probe 
electrons modifies the hybridization according to 
$H_{h}\rightarrow  (V c\dg _{0\sigma } + t_{f}p\dg_{0\sigma })f_{0\sigma }
+ {\rm H. c.}$,
where $t_{f}$ is the amplitude to tunnel directly from an $f$-state to
the probe, so that the tip modifies the orbital hybridizing with the $f$-state:
\begin{equation}\label{}
c_{0\sigma}\rightarrow c _{0\sigma } + \frac{t_{f}}{V}p_{0\sigma }.
\end{equation}
After a Schrieffer-Wolff transformation is carried out, which reduces
the Anderson model to a Kondo model, this same replacement must be
made to the Kondo interaction at site $0$ in the unperturbed Kondo Lattice model.
To leading (linear) order in $t_{f}/V$, the result of this procedure is the quoted 
result in (\ref{eq2}), where $\tilde{t}_{f}= J t_{f}/V$.

Next, we compute the tunneling current. 
One of the questions that immediately arises, is whether the
differential conductance can be analyzed in a conventional way when
cotunneling terms are present.  We now show that 
even though $\hat I$ contains a composite operator, the
weakness of the tunneling matrix elements still permits us to expand
the current  to leading order in the tunneling matrix elements, 
thereby rewriting it in terms of the full many-body
Green's functions of the bulk. To carry out this procedure, we write
steady-state tunneling current as \cite{kamenev} 
\begin{equation}\label{cur}
I (eV) \equiv \langle \hat I\rangle 
=
\frac{e}{h}{\rm Re}\int \frac{d\omega}{2 \pi}G_{p\psi }^{K} (\omega),
\end{equation}
where $G_{p\psi }^{K} (\omega)$ is the Keldysh Green
function \cite{kamenev}  between the tip electron and $\psi_{0\alpha }$. 
Expanding the current to leading order in the tunneling matrix
elements, we obtain \cite{kamenev} 
$G^{K}_{p\psi } = G_{p}^{R} G_{\psi }^{K}  + G_{p}^{K} G_{\psi}^{A}$,
where $R, A, K$ denote the retarded, advanced and Keldysh Green's
functions of the tip and the Kondo lattice. Since the tip and the lead
are in thermal equilibrium, their Keldysh Green's functions can be
re-written in terms of retarded and advanced Green's functions,
using the fluctuation dissipation relations \cite{kamenev} 
$G_{p}^{K} (\omega) = -2i \pi \rho_{p}(\omega+eV) h (\omega+eV)$
and  $G_{\psi}^{K} (\omega) = -2i \pi \rho_{\psi }(\omega) h
(\omega)$. Here $h (\omega)=1-2 f (\omega)$, where $f (\omega)$ is the Fermi distribution 
function, while $\rho_{tip}(\omega)$ is the local density of states of the tip;
$\rho_{\psi}(\omega)$ is the  ``cotunneling'' density of states of  the  sample given by  
$\rho_{\psi}(\omega)\equiv\frac{1}{\pi}\text{Im} G_\psi(\omega-i\delta),$
and $G_{\psi } (\omega)$ is the retarded Green's function
of the $\psi $ field, usually obtained through analytic continuation
of the Matsubara  imaginary time  propagator 
$G_{\psi }= - \langle T \psi_{0\alpha } (\tau )\psi_{0\alpha } \dg (0)\rangle $.

Using these relations, the current (\ref{cur}) can be re-written as
\begin{eqnarray}\label{Current}
I(eV)&=\frac{2\pi e}{\hbar}\int
d\omega~&\rho_{tip}(\omega-eV)\rho_{\psi}(\omega)
\cr&\times & (f(\omega-eV)-f(\omega)).
\end{eqnarray}
In this way, the  tunneling current into a Kondo lattice  probes the
spectral function of the composite operator.

To illustrate the tunneling into the Kondo lattice, we now solve
for the tunneling behavior in the large-$N$
limit \cite{Auerbach86,Zlatko1986,ReadNewns1987,Millis1987,Coleman1987} of the Kondo lattice, 
where $N=2j+1$ is the spin
degeneracy of the localized $f$-state. 
In this approach, the spin operator is represented 
as a bilinear of pseudo-fermions \cite{Abrikosov1965}, 
${\vec S}_f(j)=\hat{f}_{j\alpha}{\vec
S}_{\alpha\beta}\hat{f}_{j\beta}$, where ${\vec S}_{\alpha\beta}$ are
generators of the $SU(N)$ symmetry group. The mean-field theory
provides  a representation of 
the composite fermion $\left({\vec \sigma}_{\alpha\beta}\cdot{\vec
S}_f(j)\right)\hat{c}_{j\beta}$  in (\ref{eq2}) as a single fermionic
operator
\beg
\sum\limits_\beta\left({\vec \sigma}_{\alpha\beta}
\cdot{\vec S}_f(j)\right)\hat{c}_{j\beta}\to{\frac{{\cal
V}}{J}}\hat{f}_{j\alpha},
% t_f\to t_f {\cal V}/J.
\en
where the amplitude
$\frac{{\cal V}}{J}=-\langle\hat{f}_{j\beta}\dg\hat{c}_{j\beta}\rangle
$.
In this way, the large-$N$ mean field theory captures the formation of
a composite $f$-electron, an essential element of the Kondo effect.
In terms of pseudo-fermions, we can re-write single particle operator in (\ref{eq2}) as
\beg\label{eq2-2}
\hat{\psi}_{j\alpha}=t_c \hat{c}_{j\alpha}+{\tilde t}_f
\hat{ f}_{j\alpha},
\en
where the complex amplitude for tunneling into the composite 
fermion state is ${\tilde t}_f=\frac{{\cal V}}{J}t_f$.

The requirement that the number of pseudo-fermions at any
given site should be equal to $N/2$ introduces a constraint $\lambda$,  to be determined
self-consistently together with the hybridization amplitude ${\cal V}$ (see e.g. \cite{ReadNewns1987,Flint2008}). 
The resulting mean-field Hamiltonian can then be
diagonalized by means of the Bogoliubov transformation 
$\hat{c}_{\bk\sigma}=v_{\bk}\hat{a}_{\bk\sigma}+u_\bk\hat{b}_{\bk\sigma}$, 
and $\hat{f}_{\bk\sigma}=u_{\bk}\hat{a}_{\bk\sigma}-v_\bk\hat{b}_{\bk\sigma}$, where
$u_\bk$ and $v_\bk$ are the Kondo lattice coherence factors given by
$u_\bk^2=[R_\bk+(\eps_\bk-\lambda)]/2R_\bk$, $v_{\bk}^2=1-u_\bk^2$ with
$R_\bk=\sqrt{(\eps_\bk-\lambda)^2+4{\cal V}^2}$. The Hamiltonian (\ref{eq1}) in the mean-field approximation
then becomes $H_{KL}^{(mf)}=\sum_{\bk\alpha}(\omega^{-}_{\bk}\hat{a}_{\bk\alpha}\dg\hat{a}_{\bk\alpha}+
\omega^{+}_{\bk}\hat{b}_{\bk\alpha}\dg\hat{b}_{\bk\alpha})$, where $\omega^{\pm}_{\bk}=(\eps_\bk+\lambda\pm R_\bk)/2$
is the quasiparticle dispersion in the newly developed heavy Fermi
liquid. The mean-field tunneling Hamiltonian then becomes 
\beg\label{eq4}
\hat{H}_T^{(mf)}=\sum\limits_{j\alpha}\hat{p}_{j\alpha}\dg\left[t_c \hat{c}_{j\alpha}+\tilde{t}_f
\hat{f}_{j\alpha}\right]+H.c. .
\en
 Although our mean field Hamiltonian has the form of the Anderson lattice model with $U=0$, the
states on which it operates have an underlying composite structure, 
formed when local spins hybridize with conduction electrons. 
Thus, the Hamiltonian (\ref{eq4}) provides a mean-field description
of the tunneling into  the conduction band
together with the cotunneling processes involving  local moments. 

It is instructive to contrast 
the tunneling conductance expected in a Kondo lattice with that of 
a single Kondo impurity. Using the tunneling Hamiltonian (\ref{eq4}), 
we compute $G_\psi(\omega)$. In the case of a single Kondo impurity, we obtain
\beg\label{Gpsiimp}
G_\psi ^{imp}(\omega)=\frac{(t_c i\pi\rho {\cal V}+\tilde{t}_f)^2}{\omega-\lambda-i\Delta}+
t_c^2 i\pi \rho,
\en
where $\rho$ is the density of states of the conduction electrons, 
$\Delta=\pi\rho{\cal V}^2\simeq T_K$ is the  width of the Kondo resonance. The differential conductance $ \frac{dI}{dV}\equiv g(eV)$ is
\beg\label{gVsingle}
g_{imp}(eV)=N\left. 
\frac{2\pi
e^2}{\hbar}t_c^2~\rho_{tip}~\rho~
\frac{|q+\epsilon'|^2}{1+\epsilon'^2}
\right|_{\epsilon'= (eV-\lambda)/\Delta}.
\en
where $N$ is the spin degeneracy and
$q=A(eV)/B(eV)$  is the ratio of two complex tunneling amplitudes,  where
$A(eV)=\tilde{t}_f+t_c{\cal V}P(\frac{1}{eV-\epsilon_\bk})$ describes the
cotunneling
into the atomic orbital and  $B(eV)=t_c{\cal V}\pi\delta(eV-\epsilon_\bk)$ describes
direct tunneling into the metal \cite{Madhavan}. 
Here
$\epsilon'=(eV-\lambda)/\Delta$, $\rho_{tip}$ is the density of states at the Fermi level of electrons 
in the tip. 
For a broad  flat band,
$A= \tilde{t}_{f}$, $B = t_{c}{\cal V}\pi \rho  $ and $q = \tilde{t}_{f}/
(t_{c}{\cal V}\pi \rho )$. 

Now we turn to the case of the Kondo lattice. Within the large-$N$ mean field theory, we obtain
\begin{eqnarray}\label{GpsiKL}
G_\psi ^{KL}(\omega)=N\sum_\bk\frac{(t_c+\tilde{t}_f\frac{{\cal V}}{\omega-\lambda})^2}
{\omega-\epsilon_\bk-\frac{{\cal V}^2}{\omega-\lambda}},
%G_\psi ^{KL}(\omega)=\sum_\bk G_\psi ^{KL}(\bk,\omega),\\
%G_\psi ^{KL}(\bk,\omega)=\frac{(t_c+\tilde{t}_f\frac{{\cal V}}{\omega-\lambda})^2}
%{\omega-\epsilon_\bk-\frac{{\cal V}^2}{\omega-\lambda}},\nonumber
\end{eqnarray}
where $\epsilon_\bk$ is the dispersion of the
conduction band.
We obtain the following expression for the differential tunneling conductance,
\beg\label{gVKL}
\begin{split}
g(eV)=N\frac{2\pi e^2}{\hbar}t_c^2~\rho_{tip}
\sum_{s=\pm,\bk}\frac
{|q+E_{s\bk}|^2}{1+E_{s\bk}^2}
\delta(eV-\omega_{s\bk}), 
\end{split}
\en
where  
$E_{s\bk}=(\omega_{s\bk}-\lambda)/\Delta$. The prefactor of the delta-function  has a characteristic Fano functional form  \cite{Fano,Plihal2001}. This form
introduces an asymmetry in the resulting voltage dependence of the tunneling conductance $g(eV)$. 
\begin{figure}[h]
\includegraphics[width=3.5in]{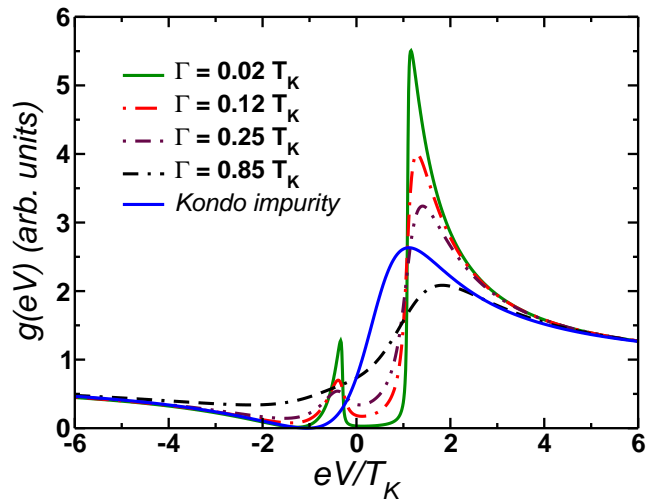}
\caption{(Color online) Differential tunneling conductance $g(V)$ for a single
Kondo impurity case (blue line) given by (\ref{gVsingle}), 
a Kondo lattice (green line) given by (\ref{gVlattice}). 
A typical Fano shape in the single Kondo impurity case
gets replaced with a double-peaked resonance line
in the Kondo lattice case.
The dashed lines illustrate the effect of  disorder, which
destroys the coherence, closing the gap in the density of states
curve. Here $\tilde{t}_f/t_c=.2$, $q
=4.9$, $\lambda/T_{K}=.3$, while
$D=100T_{K}$.
}
\end{figure}
The momentum summation in $G_\psi ^{KL}(\omega)$ (\ref{GpsiKL}) and $g(eV)$   (\ref{gVKL})
can be carried out analytically assuming a constant  conduction
electron density of states $\rho $, to give
\beg
\label{gVlattice}
g(eV)=N
\left(\frac{2 \pi e^2}{\hbar} \right)
{t}_{c}^{2}
\rho_{tip}\rho\ 
\ \frac{1}{\pi}
{\rm Im}
\tilde{G}^{KL}_{\psi } (eV-i\delta ),
\en
where 
\beg
\label{GpsiKLsummed}
\tilde{G}_\psi^{KL}(\omega)=\bigl(1
+\frac{q\Delta }{\omega-\lambda}\bigr)^2
\ln\Bigl[\frac{\omega+D_{1}-\frac{{\cal V}^2}{\omega-\lambda}}  
 {\omega-D_{2}-\frac{{\cal V}^2}{\omega-\lambda}}
 \Bigr]+\frac{2D/t_{c}^{2}}{\omega-\lambda}.
\en
Here $-D_{1}$ and $D_{2}$ are the lower and the upper  conduction band
edges respectively, and $2D=D_{1}+D_{2}$ is the bandwidth. 
%$\rho_{KL}(eV)$ is the density of states of the Kondo lattice. Within
%the mean field theory,
%\begin{equation}\label{}
%\rho_{KL}(eV)=\rho(1+\epsilon'^{-2})
%\log\Bigl[\frac{(eV+\mu)\pi\rho\epsilon'-1}
%{(eV-D)\pi\rho\epsilon'-1}
% \Bigr]+\rho(\mu+D)
% \frac{q^2(1+\epsilon'^2)}{\epsilon'(q+\epsilon')^2}
%\end{equation}
%with $\rho$ being the
% density of states of the conduction electrons.
%We assumed that $\mu+\lambda,D-\lambda\gg{\cal V}$. 
The differential tunneling conductance predicted by this formula has 
two well-pronounced peaks at $eV\sim\lambda$ separated by
a narrow hybridization gap $\Delta_g\sim 2{\cal V}^2/D$ in the single
particle spectrum, as shown in 
Fig. 2.    

In practice, experimental tunneling results will be modified by the
effects of disorder \cite{Zlatko1986}. A
phenomenological quasiparticle elastic relaxation
rate $\Gamma$ may be introduced into the theory
by replacing $\omega\rarrow \omega-i\Gamma$ in (\ref{GpsiKL}). The results of this procedure are shown
in Fig. 2. As we see, disorder removes the
sharp peak structure in the tunneling conductance $g(eV)$
(\ref{gVlattice}). The resulting lineshape of the tunneling conductance $dI/dV(eV)$ is an asymmetric smooth curve.
%This lineshape resembles the asymmetric lineshapes observed in recent experiments on Kondo lattice materials  \cite{Davis}. 

%Our results provide a basis for the interpretation of 
%scanning tunneling measurements on a Kondo lattice.  
The current work
can be extended in a number of interesting directions. 
One important aspect, is to examine the effects of
cotunneling on the fluctuations in the density of states probed in
Fourier transform STM experiments. 
In  one-band systems, the Fourier transform of these
fluctuations is phase sensitive to quasiparticle scattering \cite{Hanaguri2009, MM-PC2009},
and is expected to be an important probe of both the quasiparticle dispersion and the phase of the cotunneling
matrix elements. 
%The interplay of cotunneling
%with these processes is expected to play an important role in the
%interpretation of Fourier transformed STM. 
%phase-sensitive  Fourier-transform STM experiments, involving  observations of the quasiparticle interference.

A particularly fascinating aspect of cotunneling 
is its likely interplay with various forms of heavy fermion
order, such as heavy fermion superconductivity.  Unlike in conventional
tunneling, the quasiparticle matrix elements 
of the composite operators associated with 
cotunneling are expected to be {\sl sensitive} to the nature
of the heavy electron ground-state. 
For example, recent work has
proposed that heavy electron superconductivity may involve composite
pairing between local moments and electron pairs  \cite{Flint2008}. A key
feature of composite pairing 
is the presence of two conduction screening channels $\Gamma_1$
and $\Gamma_2$, so that now the tunneling will be described by 
the $\psi$ field (\ref{eq2}) of the form
\begin{equation}\label{}
\psi_{0\alpha }=  
t_{c}\hat{c}_{0\alpha}+\overbrace{
\sum\limits_{i=1}^2\biggl[
{t}_{f\Gamma_i}
\left({\vec \sigma}_{\alpha\beta}\cdot{\vec
S}_f(0)\right)\hat{c}_{\Gamma_i\beta}
\biggr]
}^{u\hat{f}_\alpha+v{\ \rm sgn} (\alpha )\hat{f}_{-\alpha}^\dagger},
\end{equation}
where $v$ describes hybridization in the particle-particle channel. In
this way, we see that the cotunneling term in $\psi$ may
develop both particle and hole components, resulting
in Andreev reflection even in the limit of weak tunneling.

In conclusion, we have studied electron 
tunneling into a Kondo lattice of localized moments, 
bringing out the importance of cotunneling as a primary mechanism 
of tunneling into the heavy electron fluid. We have expressed the conductance 
in terms of a spectral function of a cotunneling composite operator,
illustrating the result by a calculation carried out in the large-$N$ limit.
Our results predict that in a clean system the differential tunneling conductance 
will display two peaks separated by the hybridization gap. Addition  of disorder 
leads to the smearing of the gap and produces a Fano-like smooth asymmetric lineshape.   

The authors would like to thank R. Flint, A. Nevidomskyy, L. Greene and 
J. C. Seamus Davis for discussions related to this work. 
This research was supported by the National Science Foundation 
under Grant No. DMR 0907179.


\begin{thebibliography}{99}

\bibitem{ibm} D. Eigler, P. S. Weiss, E. K. Schweizer, and
N. D. Lang, Phys. Rev. Lett. {\bf 66}, 1189  (1991).

\bibitem{kapitulnik} C. Howald, H. Eisaki, N. Kaneko, and
A. Kapitulnik, Proc. Nat. Ac. Sci. {\bf 100}, 9705 (2003).

\bibitem{sdavis} J.E.~Hoffman, K.~McElroy, D.-H.~Lee, K.M.~Lang, H.~Eisaki,
S.~Uchida, and J.C.~Davis,   Science  \textbf{297}, 1148  (2002).

\bibitem{Hanaguri2009}  T. Hanaguri, Y. Kohsaka, M. Ono, M. Maltseva,
P. Coleman, I. Yamada, M. Azuma, M. Takano, K. Ohishi, and H. Takagi,
  Science  \textbf{323},   923  (2009).

\bibitem{Sarrao2007} see e.g. J. L. Sarrao and Joe D. Thompson, Jour. of Phys. Soc. of Jap.
{\bf 76}, 051013 (2007).

\bibitem{Coleman2008} 
P. Coleman, {\sl Handbook of Magnetism and Advanced Magnetic Materials},
Ed. H. Kronmuller and S. Parkin, John Wiley and Sons, Vol. 1, 
pp. 95-148 (2007).


\bibitem{Davis2009}  A. Schmidt, M. Hamidian, P. Wahl, F. Meier, G. Luke, and 
  J.C. Davis, Bull. Am. Phys. Soc. {\bf 54}, abstr. BAPS:2009, MAR.V29.3 (2009). 

\bibitem{Park2008} W. K. Park, J. L. Sarrao, J. D. Thompson and L. H. Greene, Phys. Rev. Lett {\bf 100}, 177001 (2008).

\bibitem{Yazdani2009} Abhay Pasupathy and Ali Yazdani, private communication (2009).

\bibitem{wyatt64} A. F. G. Wyatt, Phys. Rev. Lett. {\bf 13}, 401
(1964); R. A. Logan and J. M. Rowell, Phys. Rev. Lett. {\bf 13}, 404 (1964).

%\bibitem{rowell68}L. Y. L. Shen and J. M. Rowell, Phys. Rev. 165, 1913
%(1968).

\bibitem{Appelbaum1966} J. Appelbaum, Phys. Rev. Lett. {\bf 17}, 91 (1966).

\bibitem{Anderson1966} P. W. Anderson, Phys. Rev. Lett. {\bf 17}, 95 (1966).

\bibitem{Averin1990} D.V. Averin and Yu.V. Nazarov, Phys. Rev. Lett. {\bf 65}, 2446 (1990).

\bibitem{glazman} M. Pustilnik and L. I. Glazman, Phys. Rev. Lett. {\bf
87}, 216801 (2001). 

\bibitem{goldhabergordon}D. Goldhaber-Gordon, H. Shtrikman, D. Mahalu,
d. Abusch-Magder, U. Meirav, and M. A. Kastner, {Nature\/} {\bf 391}, 156-159 (1998).

\bibitem{vankempen}O. Yu. Koesnychenko, R. de Kort,
M. I. Katsnelson, A. I. Lichtenstein, and H. van Kempen, Nature {\bf
415}, 507-509 (2002).
\bibitem{Schrieffer1966} J. R. Schrieffer and P. A. Wolff, Phys. Rev. {\bf 149}, 491 (1966).

\bibitem{kamenev}A. Kamenev and A. Levchenko, Adv. Physics {\bf 58},
197 (2009).
%\bibitem{Mahan} G. Mahan, \emph{Many-particle physics} (Plenum Press, New York, 1990).


\bibitem{Auerbach86} A. Auerbach and K. Levin,  Phys. Rev. Lett. {\bf 57}, 877 (1986).

\bibitem{Zlatko1986} Zlatko Tesanovic, Phys. Rev. {\bf B 34}, 5212 (1986).


\bibitem{ReadNewns1987} D. M. Newns and N. Read, Adv. in Physics {\bf 36}, 799 (1987).


\bibitem{Millis1987} A. J. Millis and P. A. Lee, Phys. Rev. {\bf B
35}, 3394 (1987).

\bibitem{Coleman1987} P. Coleman,  Phys. Rev. Lett. {\bf 59},  1026 (1987).

%\bibitem{Anderson} P.W. Anderson, Phys. Rev. {\bf 124}, 41 (1961).

\bibitem{Abrikosov1965} A. A. Abrikosov, Physics (Long Island City, NY) {\bf 2}, 5 (1965).

\bibitem{Flint2008} Rebecca Flint, M. Dzero, and P. Coleman, Nature Physics {\bf 4}, 643 (2008).

\bibitem{Madhavan} V. Madhavan, W. Chen, T. Jamneala, M.F.Crommie, and N.S. Wingreen, 
Science {\bf 280}, 567 (1998).

\bibitem{Fano} U. Fano, Phys. Rev. {\bf 124}, 1866 (1961).

\bibitem{Plihal2001} M. Plihal, J. W. Gadzuk, Phys. Rev. B {\bf 63}, 085404 (2001).

\bibitem{MM-PC2009} M. Maltseva and P. Coleman,   arXiv:  0903.2752, 
accepted to Phys. \, Rev. \,B (2009).

\end{thebibliography}
\end{document}